\documentclass[twocolumn,showpacs,amsmath,amssymb,prl]{revtex4}
\usepackage{dcolumn}
\usepackage{epsfig}
\usepackage[latin1]{inputenc} 
\usepackage{bm}

\begin{document}

%
%
%
\title{Direct evidence for a characteristic dynamic lengthscale in the intermediate phase of glasses} 
\author{M. Micoulaut$^1$, M. Malki$^{2,3}$}
\affiliation{$^1$ Laboratoire de Physique Théorique de la Matière Condensée,
Université Pierre et Marie Curie, CNRS UMR 7600,  Boite 121, 4, Place Jussieu, 75252
Paris Cedex 05, France\\
$^2$ CEMHTI, CNRS UPR 3079, 1D, Avenue de la Recherche Scientifique, 45071 Orléans Cedex 02, France\\
$^3$ Université d'Orléans (Polytech' Orléans), BP 6749, 45072 Orléans Cedex 02, France}

\date{\today}
\begin{abstract}
AC conductivity spectra of xAgI-(1-x)AgPO$_3$ fast ion conducting glasses spanning the flexible, intermediate (isostatically rigid) and stressed rigid phases are analyzed. The rescaled frequency dependent spectra are mapped into time-dependent mean square displacements out of which a typical lengthscale characterizing the spatial extent $\sqrt{\langle R^2(\infty)\rangle}$ of non-random diffusion paths is computed. The latter quantity is studied as a function of AgI composition, it is found to display a maximum in the intermediate phase, providing the first clear evidence of a typical lengthscale of a dynamical nature when a system becomes isostatically rigid and enters the intermediate phase. 
\end{abstract}
\maketitle
An amorphous network progressively stiffens and becomes rigid as its connectivity or mean 
coordination number $\bar r$ increases. From a mechanical viewpoint, such an evolution can be
understood using rigidity theory which considers nearest-neighbour interactions acting at the microscopic level \cite{Traverse}, 
and enumerates the average number of constraints $n_c$ per atom and its balance with respect to the atomic degrees of freedom (3 in three dimensions). 
This has led to the recognition of a rigidity transition \cite{Phillips,Thorpe}
separating underconstrained (or flexible when $n_c<3$) from overconstrained networks (stressed rigid, when $n_c>3$).
Numerous experiments have confirmed these simple predictions \cite{Boolchand}, especially in glass science where bulk chalcogenide and oxide
glasses have been studied in detail.
\par
While the original theory predicted a single optimized glass composition where the microscopic structure is isostatic (having $n_c\simeq 3$), more recent experiments have revealed a second transition \cite{Bool1} for various network-forming glasses, providing a finite width, offering now a whole range of such isostatic compositions, and defining an intermediate phase (IP) bounded by the flexible (at low $\bar r$ or $n_c$) and stressed rigid phase (at high $\bar r$). The signature of this phase has been detected in optical \cite{Bool4}, calorimetric \cite{Bool2}, and electrical \cite{Bool3} with clear changes in behaviour when the system becomes isostatically rigid for e.g. complex heat flows at the glass transition or Raman optical elastic power-laws. Yet no typical spatial quantity giving rise to a characteristic distance (or lengthscale) has been observed in the IP, although such an isostatic lengthscale has been predicted to grow \cite{Wyart} in jammed soft sphere solids which bear some similarities to network glasses. Neutron and high energy X-ray diffraction on chalcogenide glasses have been recently reported \cite{Bitch,Billinge}, although a structural origin of the IP in typical quantities (e.g. position, width and height of the first sharp diffraction peak) related to the static structure factors was not observed.
\par
In this Letter, we show that a lengthscale associated with the intermediate phase appears in the fast ionic conductor (1-x)AgPO$_3$-xAgI which has an intermediate phase in the 9\% $<$ x $<$ 37\% AgI range \cite{agpo32}. The present quantity characterizes the spatial extent of {\em subdiffusive ion-motions}, and is computed from frequency-dependent conductivity and permittivity data together with linear response theory. The lengthscale appears to be strongly correlated with calorimetric, optical or electrical transport quantities that usually reveal the IP, but is neither associated directly with static network structure nor with changes in molar volume, as both evolve smoothly with glass composition. Another lengthscale characterising the typical distance mobile ions travel to overcome backward-forward driving forces is found to be only sensitive to the onset of network flexibility at x=37\% AgI \cite{agpo32}. The present findings highlight, therefore, the clear dynamical nature of the lengthscale underlying the IP, and underscores a possible connection with the peculiar relaxational phenomena of isostatic glass-forming liquids \cite{Micoul}.
\par
Details on the synthesis and the structural, calorimetric, electrical and spectroscopic properties of dry and homogeneous bulk (1-x)AgPO$_3$-xAgI glasses were discussed earlier in Refs. \cite{agpo3}-\cite{agpo3f}.
Conductivity measurements were performed on disks about 10 mm in diameter and 2 mm thick on which Pt electrodes were deposited. The complex impedance Z$^*$($\omega$)=Z'($\omega$)+iZ$^{''}$($\omega$) was measured by a Solartron SI 1260 impedance analyzer over the frequency range of 1 Hz-1 MHz. Three temperatures have been considered: 250 K, 300 K and 350 K. Conductivity $\sigma^*$($\omega$) was deduced from the complex impedance Z$^*(\omega)$, sample thickness, t, and surface area S covered by platinum using the formula: $\sigma^*$=t/S$\times$1/Z$^*$. The real part of the dielectric permittivity $\varepsilon'(\omega)$ was computed using the electrical modulus M$^*$: M$^*$=1/$\varepsilon^*$=i$\omega\varepsilon_0/\sigma^*$, where $\varepsilon_0$ is the permittivity of free space. Because of electrode polarisation occuring at higher temperatures (see Fig. 1c and text below), we have restricted the analysis to the temperature of 250 K.

\begin{figure}
\begin{center}
\epsfig{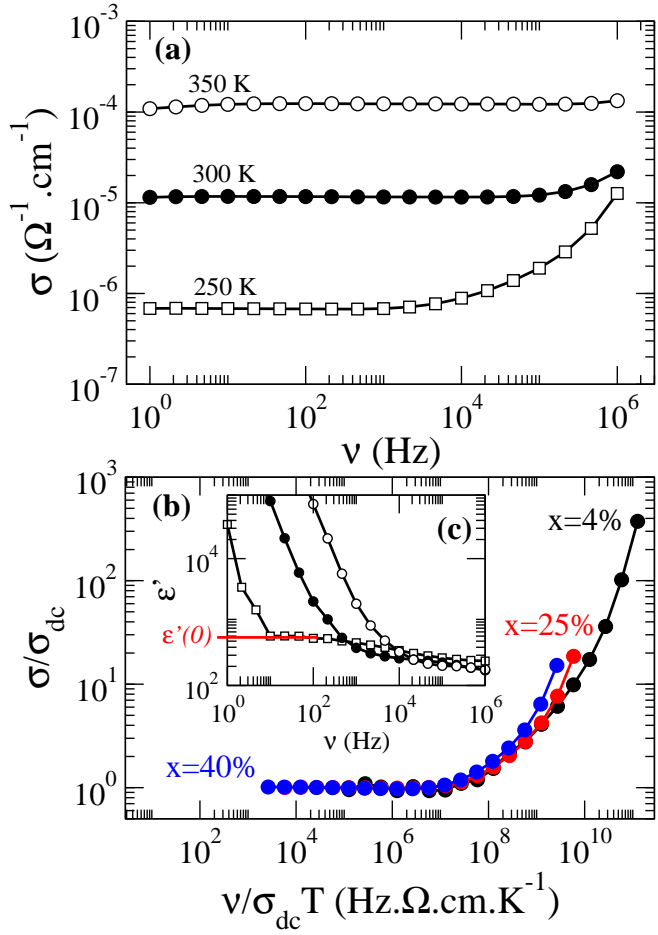}
\end{center}
\caption{(color online)\label{fig1} (a) Experimental conductivities of a 25AgI-75AgPO$_3$ glass at three different temperatures and corresponding (c) dielectric permittivities. The red horizontal line serves to define the low frequency permittivity $\varepsilon'(0)$ used in equ. (\ref{epsilo}). (b) Master curve $\sigma/ \sigma_{dc}$ as a function of the rescaled frequency $\nu/\sigma_{dc}T$ for three selected compositions at T=250 K, derived from the experimental conductivities.}
\end{figure}
\par
Figure \ref{fig1}a shows the log-log plot of the real part of the conductivity with frequency at three different temperatures for a glass at x=25\% AgI. Dielectric permittivity spectra $\varepsilon'(\omega)$ were computed (Fig. \ref{fig1}c) and tracked with composition \cite{Bool3}. Concerning the latter, electrode polarization effect onsets at low frequency manifested by a rapid growth of $\varepsilon'$. This phenomenon hides the intrinsic frequency behaviour of $\varepsilon'(\omega)$ so that an evaluation of the limit $\varepsilon'(0)$ (defined in Fig 1c) is only possible at low temperature \cite{Ngai}. Similar spectra and features were obtained at other compositions and temperatures (see Ref.\cite{Bool3}). At low frequency and low temperature (250 K), we obtain a dc régime which  displays an Arrhenius behaviour, whereas at higher frequencies a dispersive régime is observed, with $\sigma$ or $\varepsilon'$ becoming frequency dependent. A frequency $\nu_p$(x,T) (also called dielectric loss peak frequency) satisfying $\sigma(\nu_p)/\sigma_{dc}$=2 is usually introduced to characterize the cross-over range between the dc and dispersive conductivity régimes, the latter being associated with a subdiffusive régime in the time domain valid for t$<$t$_p$=1/(2$\pi\nu_p$). The frequency $\nu_p$ increases with increasing temperature, and a simple rescaling by a factor $\sigma_{dc}T$ has been proposed \cite{Rol1} in order to produce a master-curve for various temperatures and compositions. Results are shown in Fig. 1b for three selected compositions which show indeed that the data nearly map onto the same curve.
\par
\begin{figure}
\begin{center}
\epsfig{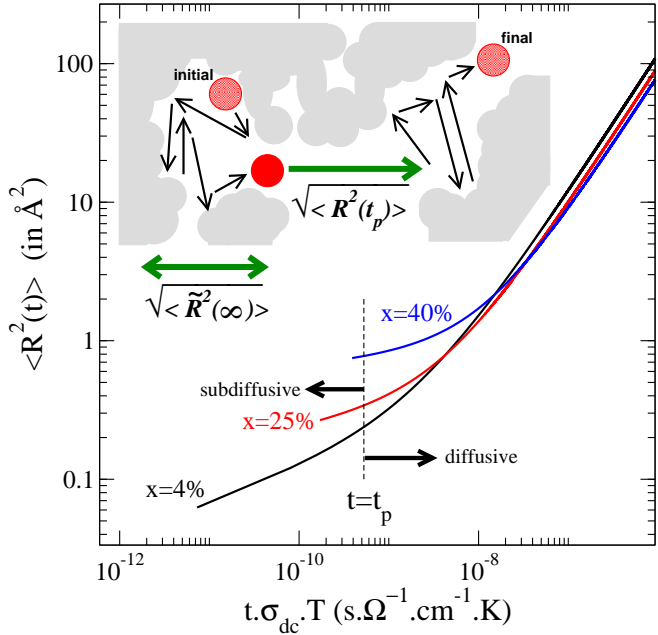}
\end{center}
\caption{(color online) \label{msd1} Mean-square displacement $\langle R^2(t)\rangle$ at 250 K of (1-x)AgPO$_3$-xAgI at three selected compositions representing the three phases of interest, derived from the master curve represented in Fig. 1 and equ. (1) as a function of rescaled time t$\sigma_{dc}$T. The vertical broken line correspond to the abcissa t$_p\sigma_{dc}T$ defining t$_p$ for a glass at x=4\%, the approximate limit between the diffusive and sub-diffusive régimes. Inset: Trajectory of an ion between an initial and a final state, which serves to define the quantities $\sqrt{\langle \tilde R^2(\infty)\rangle}$ and $\sqrt{\langle R^2(t_p)\rangle}$.}
\end{figure}

We now build on the linear response theory developed by Roling and co-workers \cite{Roling1,Roling2} and recently applied to borophosphate glasses \cite{Eckert}. It allows extracting from conductivity spectra a mean square displacement $\langle r^2(t)\rangle$ of the mobile ions given by: 
\begin{eqnarray}
\langle r^2(t)\rangle&=&{\frac {12k_BTH_R}{N_vq^2\pi}}\int_0^t dt'\int_0^\infty {\frac {\sigma(\nu)}{\nu}}\sin(2\pi\nu t')d\nu \nonumber \\ 
&=&\langle R^2(t)\rangle H_R.
\end{eqnarray}
$N_v$ is the charge density, and $\langle R^2(t)\rangle$ is the mean-square displacement of the center of charge of the mobile ions. $H_R$ is the so-called Haven ratio \cite{Roling2}, found usually between 0.2 and 1.0 and depends on the concentration of charge carriers \cite{Sidebott,Greaves}. It characterizes the degree of cooperatvity of ion motion and can only be obtained by combining measurement of $\sigma_{dc}$ with the low-frequency limit of the self-diffusion constant $D'(0)$. As $H_R$ is not known for the present system, we focus on $\langle R^2(t)\rangle$. 
From the permittivity spectra, one can extract \cite{Roling1,Roling2} the long-time limit of a rescaled quantity, $\langle R^2(\infty)\rangle$, using $\langle R^2(t)\rangle$, given by:
\begin{eqnarray}
\label{epsilo}
\langle \tilde R^2(\infty)\rangle&=&lim_{t\rightarrow\infty} {\frac {1}{6t}}\biggl[\langle R^2(t)\rangle-6D'(0)t\biggr] \nonumber \\
&=&{\frac {6k_BT\epsilon_0}{N_vq^2}}\biggl[\varepsilon'(0)-\varepsilon'(\infty)\biggr]
\end{eqnarray}
Figure 2 shows $\langle R^2(t)\rangle$ for three selected compositions corresponding to the flexible (40\% AgI), intermediate (25\%) and stressed rigid phase (4\%), computed from equ.(1). At long time (t$>$t$_p$), $\langle R^2(t)\rangle$ displays the diffusive régime as detected from the slope of 1 in the log-log plot, while a sub-diffusive régime corresponding to correlated forward-backward motions appears at shorter times scales (t$<$t$_p$). Here the limit $t_p$ (or its rescaled quantity $t_p\sigma_{dc}T$) designates the time beyond which ions start to diffuse. One can therefore consider the characteristic length $\sqrt{\langle R^2(t_p)\rangle}$ to be the typical distance mobile ions have to travel to overcome the interactions responsible for these forward-backward motions (see Fig. \ref{msd1} inset). 
According to the definition of the self-diffusion coefficient, for mobile ions the quantity $\langle r^2(t)\rangle-6D'(0)t$ vanishes at the (diffusive) long-time limit. This property is not satisfied for the related $\langle R^2(t)\rangle$ characterizing the motion of the center of charge of the mobile ions, and the quantity $\langle \tilde R^2(\infty)\rangle$ given by equ. (2) provides a measure of the spatial extent (Fig. 2, inset) of sub-diffusive motions \cite{Roling2}.
\par
Both lengthscales $\sqrt{\langle R^2(t_p)\rangle}$ and $\sqrt{\langle \tilde R^2(\infty)\rangle}$ can be determined from the conductivity and permittivity spectra, and we follow these with AgI content of glasses. Results are displayed in Fig. 3. $\sqrt{\langle \tilde R^2(\infty)\rangle}$ displays a maximum in the IP, driven mainly by a maximum in the permittivity difference $\varepsilon'(0)-\varepsilon'(\infty)$ appearing in equ.(2) \cite{Bool3}. Starting from 5.5 $\AA$ at x=0, this lengthscale increases up to a maximum of nearly 9 $\AA$ in the IP and then decreases to 2.5 $\AA$ in the flexible phase. Trends in $\sqrt{\langle \tilde R^2(\infty)\rangle}$ as a function of x clearly correlate with those in the non-reversing heat flow, $\Delta H_{nr}$, obtained from calorimetric measurements \cite{agpo32}, and with the three régimes for dc conductivity. These trends cannot be explained on the basis of free volume changes induced by AgI doping as molar volumes are known \cite{agpo3f} to decrease linearly. 
These results suggest that there is a maximum in spatial extent of subdiffusive motions in the IP. At low AgI concentrations, isolated subdiffusive regions (SDR) are embedded in a stressed rigid structure and lead to a low spatial extent. The increase of the number of hopping sites in the IP \cite{Bool3} leads to a growth of the subdiffusive regions (and $\sqrt{\langle \tilde R^2(\infty)\rangle}$). Finally, once these SDR percolate, as also suggested from Molecular Dynamics simulations \cite{MD}, the spatial extent of these regions decreases. Fig. 3 illustrates for the first time that a typical lengthscale is associated with the IP, which is of dynamical origin. There is no structural origin of such a lengthscale because relevant quantities deduced from X-ray diffraction \cite{Xray} behave smoothly with composition, an observation also reported on a different systems in \cite{Billinge} .
\par
\begin{figure}
\begin{center}
\epsfig{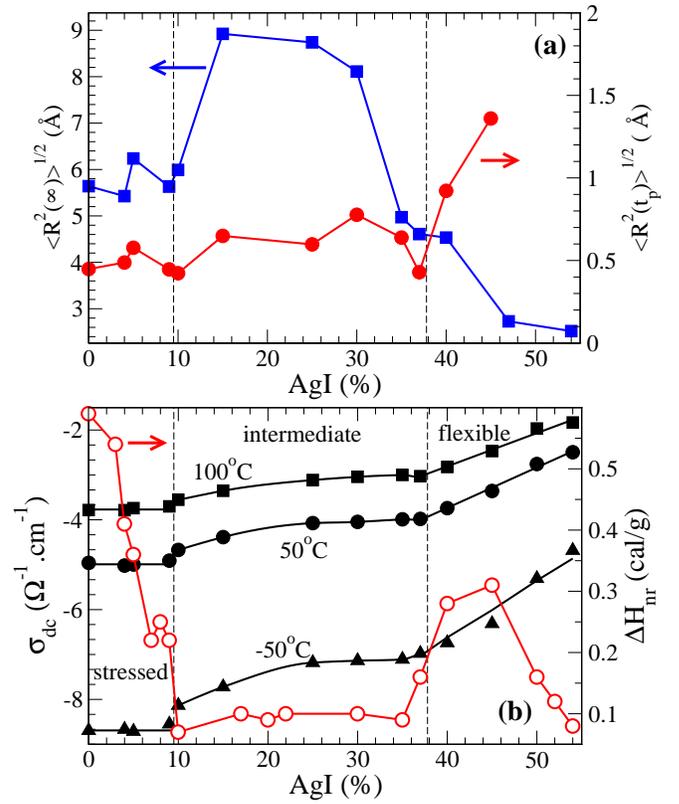}
\end{center}
\caption{(color online) (a) Plot of the characteristic lengthscales $\sqrt{\langle \tilde R^2(\infty)\rangle}$ (blue) and $\sqrt{\langle R^2(t_p)\rangle}$ (red, right axis) in (1-x)AgPO$_3$-xAgI as a function of AgI composition, compared to (b) (extracted from Ref. \cite{Bool3}): conductivity $\sigma_{dc}$ (black filled symbols) at different temperatures and non-reversing heat flow $\Delta H_{nr}$ (open red circles, right axis). The vertical broken lines define the intermediate phase boundaries.}
\end{figure}
The other quantity of interest, the lengthscale $\sqrt{\langle R^2(t_p)\rangle}$, is found to display a threshold at the intermediate to flexible transition (Fig. 3, right axis), also known as the rigidity transition. In fact, as long as the system is rigid (stressed or isostatic), this lengthscale remains nearly constant at about 0.5 $\AA$, largely because a large energy is needed to locally deform the network in order to overcome the backward driving forces causing the correlated subdiffusive forward-backward motions. One control parameter of $\sqrt{\langle R^2(t_p)\rangle}$, the time t$_p$ for the onset of diffusion decreases from 400 $\mu$s at x=0 to 10-15 $\mu$s close to the flexible transition (37-40\%). Once the system has become flexible and floppy mode proliferate, ion motion is facilitated and the time decreases to t$_p$=6.7 $\mu$s at x=45\% AgI. At this concentration, and because of the large number of carriers, the time needed to leave a SDR reduces, but a larger distance is needed to overcome the backward driving forces induced by the presence of an increased number of Ag cations.  
\par
In summary, we have shown for the first time that a typical lengthscale, of a dynamical origin, is associated with the Intermediate Phase. This lengthscale appears in fast-ion conducting glasses, and measures the spatial extent of subdiffusive ionic motions that increase when the system becomes isostatic and decreases once conduction pathways percolate in the flexible phase. Another lengthscale characterizing the forward-backward motion at play in the subdiffusive régime is found to signal only the intermediate to flexible transition when the network softens and leads to a strong increase of the dc conductivity. These results found for a fast ion conducting glass may also exist in covalent network glasses. Anomalies in viscosity have been observed in network glass-forming liquids such as binary Ge-Se \cite{visco}, and since viscosity and diffusion are related via the classical Nernst-Einstein relation, such a dynamical lengthscale may exist in e.g. Ge and Se diffusion measurements. 
\par
It is a pleasure to acknowledge many stimulating discussions with P. Boolchand, B. Goodman, J.C. Phillips, P. Simon. This work is supported by Agence Nationale de la Recherche (ANR) n.09-BLAN-0190-01.

\end{document}